\documentclass{IET-Conf-Paper}

\header{9\textsuperscript{th} E-Mobility Power System Integration Symposium | Berlin, Germany | 06-07 October 2025}

\usepackage{pgfgantt}
\usepackage{xcolor}
\usepackage{tikz}
\usetikzlibrary{arrows.meta, positioning}
\usepackage{tabularx}  
\usepackage{float}  
\usepackage{placeins}
\usepackage[font=small, labelfont=bf, justification=centering, format=plain]{caption}
\usepackage{parskip}
\usepackage{array} 
\newcolumntype{L}[1]{>{\raggedright\arraybackslash}p{#1}}
\usepackage[justification=raggedright,singlelinecheck=false]{caption}
\usepackage{needspace}

\begin{document}

\title{Enabling Intelligent Bidirectional Charging: A Real-World Communication Interface Between Electric Vehicles, Charging Infrastructure, and a Control Optimizer}

\author{
Shangqing Wang\ad{1}, 
Abhirup Sain\ad{1},
Christopher Lehmann\ad{1},
Shiwei Shen\ad{1},
Razan Habeeb\ad{1},
Frank H. P. Fitzek\ad{1,2}\corr
}

\address{
\add{1}{Deutsche Telekom Chair of Communication Networks, TU Dresden, Dresden, Germany}  
\add{2}{Centre for Tactile Internet with Human-in-the-Loop (CeTI), TU Dresden, Dresden, Germany}  
\add{*}{shangqing.wang@tu-dresden.de, abhirup.sain@mailbox.tu-dresden.de, 
christopher.lehmann@tu-dresden.de, shiwei.shen@tu-dresden.de, razan.habeeb@tu-dresden.de, frank.fitzek@tu-dresden.de}
}


\begin{abstract}
This paper presents the real-world implementation and field validation of a user-aware, smart bidirectional electric vehicle (EV) charging system, developed for the Mobilities for EU and DymoBat projects in Dresden. Building on previous scenario-based modeling and simulation frameworks, the system takes a critical next step: enabling the transition from conceptual pilots to operational deployment in city contexts. To support increasing demands for grid flexibility and clean urban mobility, our solution couples real-time user and vehicle telemetry with a centralized optimization platform, achieving data-driven charging and discharging decisions. The architecture integrates a wireless On-Board Diagnostic II (OBD-II) interface and an open communication middleware node with a 5G campus network to furnish early, pre-connection access to vehicle state-of-charge, even before the vehicle is plugged in. Simultaneously, a tablet-based human-machine interface captures user preferences—such as desired departure time and energy needs—which inform optimization logic alongside grid conditions and mobility trends.

A key innovation is the closed-loop, multi-level communication architecture that links the user, the EV, the bidirectional charging station, and the grid control center, leveraging the Open Charge Point Protocol (OCPP) for seamless system integration. This design synthesizes software, embedded hardware, and networked components to deliver real-time, driver-inclusive charging management. Field deployment at Ostra Sport Park, located in the Dresden Ostra District, confirms the operational feasibility of the system, demonstrating improved load balancing and robust vehicle-to-grid (V2G) operation in practice. By enabling early data acquisition and proactive system management, the framework sets a benchmark for positive energy districts (PEDs) and climate-neutral city initiatives across Europe. This work represents the core technical result of Dresden’s pilot for the Mobilities for EU project, advancing the applied engineering and user-centered integration needed for the next generation of urban e-mobility systems.
\end{abstract}

\maketitle
\section{Introduction}

The integration of electric vehicles (EVs) into urban transportation infrastructures presents transformative opportunities for cities driving toward sustainability and energy efficiency~\cite{prata2013moving}. Recent advances, including smart and bidirectional charging and vehicle-to-grid (V2G) strategies, have proven technical potential to reduce grid stress, enable renewable integration, and enhance operational flexibility for urban energy systems~\cite{wang2024b_usecase}

Building on our earlier research in Dresden’s Ostra District, where simulation architectures were developed to evaluate charging strategies and facilitate stakeholder engagement~\cite{Wang2025:Simulationb}, this work takes the next crucial step: translating simulation-driven insights into practice through a real-world pilot deployment. Within the framework of the Mobilities for EU project, we focus on bridging the persistent gap between modeled technical feasibility and real-world implementation. In particular, we address the need for integrated communication and control architectures that enable seamless data flow between vehicles, users, and charging infrastructure, regardless of whether or not an EV is physically connected to a charger.

This paper presents a high-level system architecture and technology transfer pathway for a flexible EV charging ecosystem that links driver preferences, live EV telemetry, and centralized optimization logic. Our approach leverages a wireless On-Board Diagnostic II (OBD-II) interface (utilizing the HEX-NET Wi-Fi module with VCDS), a lightweight computing node for local data processing and secure relay, and 5G connectivity to transmit real-time vehicle data to a central control center. The backend utilizes the DymoBat optimization platform, developed by our team for advanced scheduling of smart charging and V2G operations~\cite{dymobat2025}. In parallel, a lightweight mobile application allows drivers to specify preferences and mobility intentions, closing the loop for user-centric, predictive charging control.


Unlike conventional systems that only gather data once a vehicle is plugged in, our architecture enables early awareness of vehicle status and user intent, allowing for proactive resource management of EVs even while parked but not yet connected. In fact, operational data capture can begin even earlier: drivers are able to reserve charging stations and specify charging preferences remotely via the mobile application before arrival, enabling the system to anticipate demand and optimize scheduling ahead of time. This early and continuous data access is foundational to anticipatory grid services and supports more adaptive, user-centric operational strategies.

Our proposed concept forms the basis for an upcoming large-scale deployment in Dresden’s Ostra District, set to begin following the installation of bidirectional charging stations in late 2025. By addressing both the practical and architectural requirements for integrated system operation, this work marks a key step from simulation-based evaluation to field-ready, scalable solutions for smart urban mobility.

\section{Background and Motivation}

The integration of EVs with advanced charging infrastructure is recognized as one of the essential elements of the sustainable transformation of urban energy and mobility systems~\cite{micari2024electric, SUMP2025plus, CIVITAS2017}. In recent years, significant progress has been made in modeling and simulating the potential of smart charging, bidirectional charging, and V2G strategies. Modular simulation environments and co-simulation frameworks have enabled researchers and planners to analyze operational, regulatory, and business impacts at city scale~\cite{rehman2019cloud, IEEE_HLA_2010, Wang2025:Simulationb, wang2024a5G, MobilitiesForEU}.

Our prior studies in Dresden’s Ostra District have shown that advanced charging logic can reduce peak demand, improve energy efficiency, and provide a valuable decision-support tool for stakeholders~\cite{Wang2025:Simulationb, wang2024a5G, wang2024b_usecase}. However, as recognized in literature and our own experience~\cite{haumer1999bridging, delRioCarazo2022, Wang2025:PDTT}, simulation remains primarily a preparatory phase; its insights must ultimately be validated and realized through operational deployment. However, persistent gaps remain in bridging simulation and operational practice, particularly in the timely and actionable integration of vehicle and user data for real-world charging and grid optimization.

A critical gap identified both in our current testbed and across most documented bidirectional charging pilots is the lack of timely, granular information about vehicle status, such as State of Charge (SoC) and user preferences, prior to the vehicle’s physical connection to a charging station. Systems, including our own campus-based research platform, typically access such data only when the vehicle is plugged in, leaving substantial distributed energy potential unaccounted for and limiting the ability to balance grid resources or deliver user-centric services proactively. While proprietary or experimental solutions (e.g., some telematics APIs) may offer partial pre-connection insight~\cite{Schellenberg2021telematics, Nordholm2023EVreview}, open and scalable implementations remain rare~\cite{SciurusFinalReport, ParkerProject, Wang2025:PDTT}. As the Mobilities for EU project prepares for city-scale deployment in the Ostra District, the need for an architecture that overcomes these limitations is clear~\cite{MobilitiesForEU}.

\section{System Requirements and Design Objectives}

The transition from simulation-based planning to real-world deployment requires system architectures that deliver tangible operational functionality beyond scenario analysis. Prior work, including simulation studies in Dresden’s Ostra District, has established the potential benefits of smart and bidirectional charging~\cite{Wang2025:Simulationb}; however, most existing implementations continue to rely on plug-in events to trigger data collection and optimization~\cite{wu2018transactive}. This constraint limits their utility for proactive grid management and the realization of user-oriented charging strategies.

Our objective is to enable early and consistent access to both vehicle and user data, independent of charging status, using scalable and commercially available components. This section outlines the core requirements and design objectives established to directly address the identified operational and technological gaps.

\subsection{Core Objectives of the Architecture}

\begin{itemize}
    \item \textbf{Early Access to Critical Data:} Enable the automated, real-time collection of vehicle SoC, location, and user mobility preferences before the vehicle is physically connected to a charger, supporting predictive energy management and resource forecasting. By incorporating advanced registration and ongoing location tracking, the system allows city planners to anticipate exactly when and where bidirectional charging resources will be available, enabling optimized grid operations and infrastructure utilization.
    
    \item \textbf{Seamless, Standardized Communication:} Establish robust, bidirectional communication using wireless OBD-II interfaces, a lightweight processing unit in-car, mobile apps, and private 5G campus networks, to ensure interoperability, security, and scalability.
    
    \item \textbf{Deployability and Scalability:} Leverage commercial off-the-shelf hardware and modular software to facilitate rapid replication and integration across urban environments of differing scales.
    
    \item \textbf{Urban System Adaptability:} Design architectural flexibility to accommodate diverse regulatory, technical, and planning contexts, making the solution agnostic to hardware generations or urban platform specifics.
\end{itemize}

\subsection{Comparative Assessment of System Requirements}

Table~\ref{tab:requirements} highlights how our approach addresses the current constraints of conventional plug-in-dependent systems.

\begin{table*}[!t]
\centering
\caption{Feature Comparison of EV Systems}
\label{tab:requirements}
\renewcommand{\arraystretch}{1.2}
\begin{tabular}{|L{4.2cm}|L{6.3cm}|L{6.3cm}|}
\hline
\textbf{Requirement} & \textbf{Proposed Architecture} & \textbf{Conventional Systems} \\
\hline
Pre-connection data access & SoC and driver intentions gathered wirelessly before plug-in & Data only available after EV connects to charger \\
\hline
User preference integration & Mobile app enables real-time and scenario-based preference input & Lack of dynamic user input; defaults or static preferences \\
\hline
Platform and hardware agnosticism & Commercial OBD-II, in-car processing unit, and app-based interfaces enable modular rollout & Typically vendor-specific, closed systems with limited adaptability \\
\hline
Scalability and replication & Designed for lightweight, city-scale deployment using standard 5G networks and open protocols & High-cost infrastructure and tight coupling between vehicle and station \\
\hline
Urban flexibility & Architecture supports regulatory variation and integration with city-specific mobility ecosystems & Systems often lock users into predefined charging environments \\
\hline
\end{tabular}
\end{table*}

Grounding each requirement in operational needs, this architecture provides a real-world-ready blueprint for city-scale, distributed EV-grid integration. These requirements inform the design of a modular, communication-first architecture: a system in which real-time, bidirectional communication between users, vehicles, charging stations, and grid services is not an add-on, but a foundational design principle. By prioritizing early data exchange and protocol-driven interoperability, this architecture enables anticipatory and user-aware control logic, addressing scalability, flexibility, and integration challenges in modern urban mobility contexts. The next section details the high-level system architecture and component interactions enabling this vision.

\section{Proposed High-Level Architecture}
\begin{figure}[h]
\begingroup
\setlength{\parskip}{0pt} 
\setlength{\belowcaptionskip}{-10pt} 
\centering
\includegraphics[width=\linewidth]{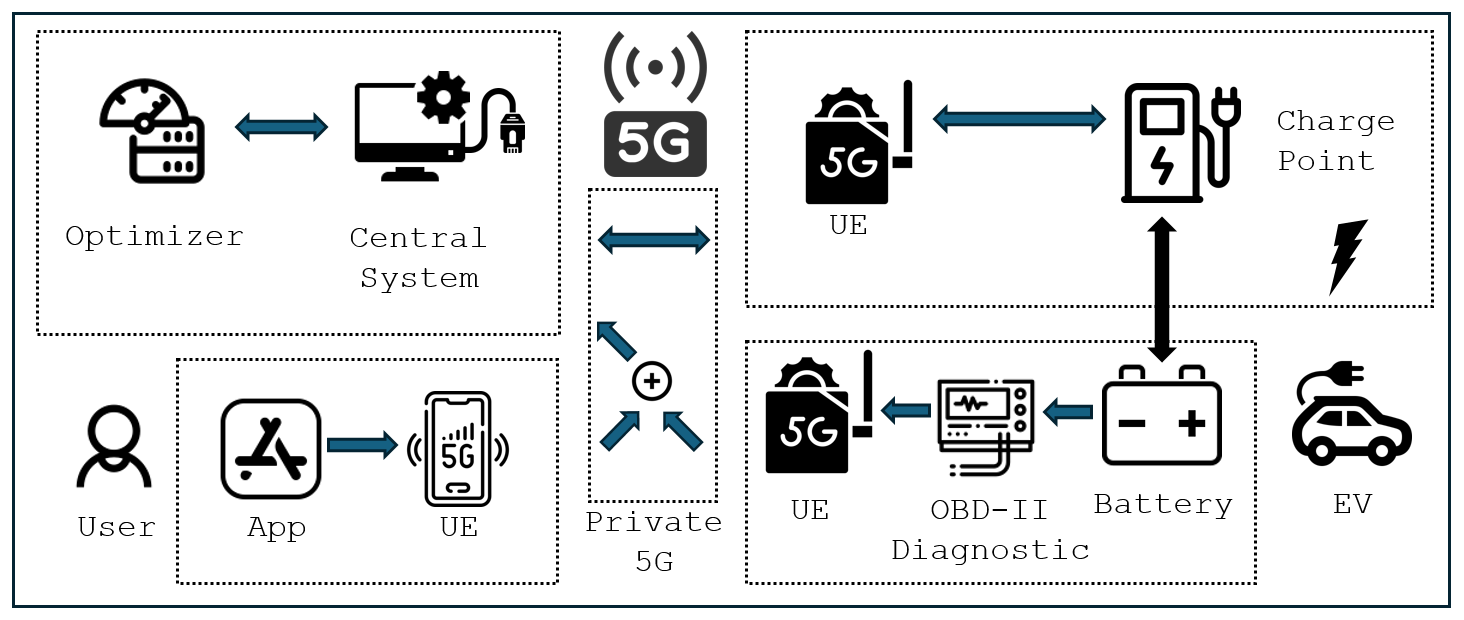}
\caption{High-Level Architecture: System Overview. The diagram shows key system elements: OBD-II interface, in-car processing unit, a private 5G network inclindng User Equipments (UEs), central optimizer, and charge point, enabling closed-loop data exchange between vehicle, user, and grid.}
\label{fig:systemoverview}
\endgroup
\end{figure}

This section presents the high-level system architecture, as shown in Fig.~\ref{fig:systemoverview}, designed to enable a user-aware, bidirectional EV charging ecosystem in an urban context. The architecture is modular and communication-centric, integrating real-time vehicle telemetry, user preference capture, edge data processing, and centralized optimization via a secure 5G network. While components such as the wireless OBD-II interface, in-car processing unit including a 5G communication relay, and driver mobile app are under active development, core elements like the private 5G network and backend optimizer are already operational. Collectively, these modules facilitate seamless data exchange and anticipatory control, forming the foundation for scalable deployment and field validation.

\Needspace{6\baselineskip} 
\textbf{System components:}

\subsection{Wireless OBD-II interface on the EV:}
The wireless OBD-II interface enables automatic, real-time retrieval of SoC, battery capacity, and diagnostic data directly from the EV. Unlike conventional plug-in approaches, this module grants the backend access to vehicle status prior to physical connection, supporting anticipatory scheduling and more accurate prediction of charging requirements. Data is transmitted securely via wireless link from diagnostic device to the in-car processing unit, with strong emphasis on reliable connectivity, encryption, and authentication to protect sensitive vehicle information. This interface forms the foundation for proactive grid interactions within the bidirectional charging ecosystem.

\subsection{In-Car Processing Unit:} 
The processing unit serves as a secure in-car relay, connecting the private 5G network, the driver mobile app, and the in-car OBD-II interface. It lowers backend load and guarantees data integrity prior to sending it to the optimizer by combining and pre-processing telemetry (such as SoC, capacity, and user preferences). Distributed intelligence is supported by this edge-layer design, which allows for scalable deployment across several EVs while preserving modularity for upcoming interfaces or services. To protect sensitive vehicle and user data, security measures like encrypted communication and device authentication are given top priority. The in-car processing unit serves this function by providing flexibility for local logic, quick prototyping, and smooth integration into the bidirectional charging ecosystem in addition to relaying information.

\subsection{Private 5G Network:} 
\begin{figure}[h]
\centering
\includegraphics[width=1.0\linewidth]
{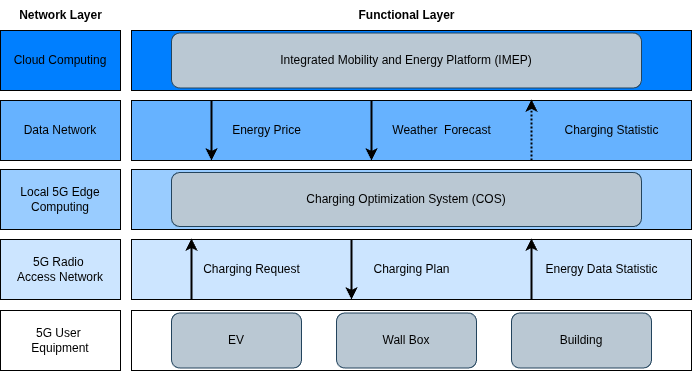}
\caption{Network Layer and Functional Layer Overview. The architecture features a private 5G network comprising 5G radio access network (RAN), local 5G edge cloud infrastructure, and separate functional entities for data aggregation, optimization, and charging control.}
\label{fig:5g-functional-layer}
\end{figure}

The dedicated private 5G campus network forms the secure, high-performance communications backbone for the entire system. For V2G integration, it provides the ultra-low latency, reliability, and data security required for real-time information exchange among vehicles, drivers, charging stations, and central controllers. Unlike public mobile or Wi-Fi networks, the isolated setup offers guaranteed quality of service (QoS) and operates on licensed local spectrum to prevent interference, enabling precise network slicing to isolate critical charging and mobility data. The architecture comprises a local radio access network and edge-hosted 5G core, consistently achieving sub-15 ms round-trip latency~\cite{wang2024a5G}. As depicted in Fig.~\ref{fig:5g-functional-layer}, this design supports coordinated, scalable, and secure operations across concurrent mobility applications.

\subsection{Driver mobile app/interface for preference input:} 
\begin{figure}[H]
\begingroup
\setlength{\parskip}{0pt} 
\setlength{\belowcaptionskip}{-10pt} 
\centering
\includegraphics[width=0.9\linewidth]{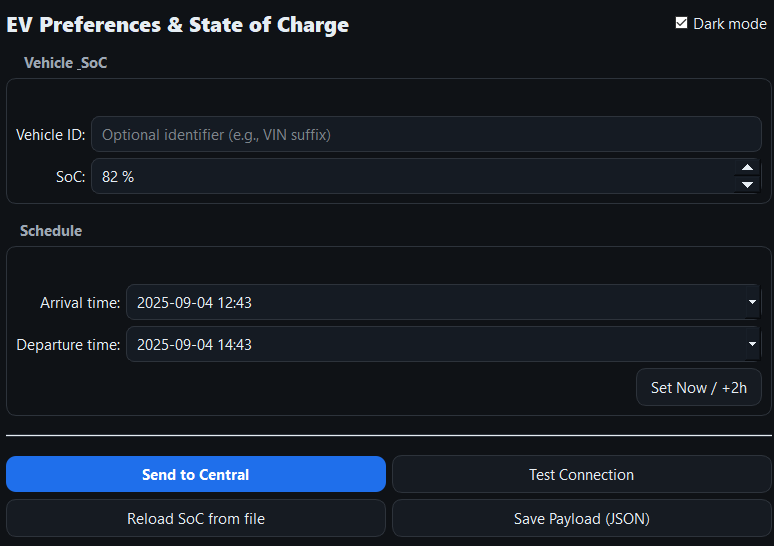}
\caption{Driver Mobile App Prototype. The application captures essential user inputs such as vehicle SoC, arrival and departure times, and securely transmits them via the in-car processing unit and 5G network to the backend optimizer.}
\label{fig:app}
\endgroup
\end{figure}

A cross-platform mobile application is under development to capture driver preferences in real time as depicted in Fig.~\ref{fig:app} including intended departure times, required energy, and planned mobility. By enabling the system to react to actual user needs instead of relying on predetermined defaults, this interface makes the architecture fully user-aware. Preferences are securely transmitted via the in-car processing unit and 5G network to the backend optimizer, where they are fused with live grid and vehicle data to compute optimal charging strategies. Beyond data capture, the app delivers clear charging status feedback and strategy recommendations to the driver, fostering engagement and trust. Interactive prototypes have already undergone early usability testing, and the development process is focused on continual improvement and user experience prior to full deployment.

\subsection{Centralized Optimizer/Control Center (Backend)}
The backend optimizer, implemented via the DymoBat platform, centrally manages energy scheduling and system optimization. It ingests real-time vehicle telemetry, driver preferences, grid conditions, and operational constraints to generate scenario-based charging and discharging schedules that balance user needs, grid flexibility, and renewable integration. Optimization decisions are communicated back over the 5G network for execution by local charging infrastructure and, where relevant, feedback to drivers through the mobile app.

\begin{figure}[b]
\centering
\includegraphics[width=0.9\linewidth]{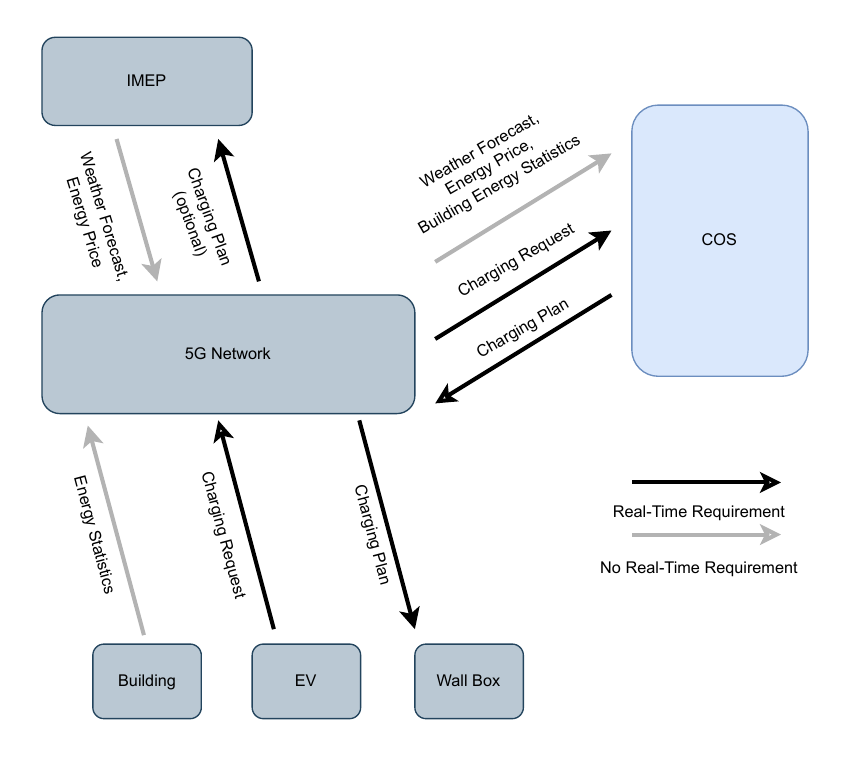}
\caption{Urban Implementation System Architecture. Transition from campus testbed direct communication to a scalable 5G-enabled platform incorporating the Integrated Mobility and Energy Platform (IMEP) for data aggregation and Charging Optimization System (COS) for local control. This architecture enables simultaneous, low-latency communication with multiple EVs, charging stations, and the centralized backend optimizer.}
\label{fig:urban-system-arch}
\end{figure}

As depicted in Fig.~\ref{fig:urban-system-arch}, two principal components support this centralized logic: the Integrated Mobility and Energy Platform (IMEP), which aggregates and processes heterogeneous data streams such as vehicle status, energy prices, and environmental conditions; and the Charging Optimization System (COS), which translates global optimization outputs into actionable, site-specific charging commands. The modular integration of these components enables scalable and adaptive control across multiple devices and deployment sites. Ongoing collaboration with the DymoBat development team aims to further enhance real-time field data integration and responsiveness.

\subsection{Data Flow and Communication Protocols}
The architecture enables closed-loop, multi-level data flow connecting EVs, drivers, charging assets, and the optimization backend. Vehicle state and driver preferences are collected locally (wireless OBD-II, app), relayed by the in-car processing unit over the 5G network, and processed centrally. Charging schedules and operational commands are then sent back to the relevant field devices or interfaces.

For secure and standardized communication, the following protocols are implemented or planned:
\begin{itemize}
 \item \textbf{ISO 15765-4 (Controller Area Network (CAN) bus over OBD-II)}: Defines the in-vehicle communication protocol used by most modern EVs to exchange diagnostic and telemetry data via the OBD-II interface. This enables reliable retrieval of parameters such as State of Charge, battery voltage, and current, providing the backend with early pre-connection access to critical vehicle status.~\cite{ISO15765-4}
    \item \textbf{ISO 15118-20: Protocol requirements for Vehicle-to-Grid Communication} Ensures secure and interoperable bidirectional charging communication directly between EVs and charging stations (wallboxes)~\cite{ISO15118-20}
    \item \textbf{IEC 63584: Open Charge Point Protocol (OCPP):} Manages charging station operations and integration with backend systems~\cite{OCPP}.
    \item \textbf{Message Queuing Telemetry Transport (MQTT)}: A lightweight, asynchronous publish/subscribe messaging protocol used for efficient and reliable data transfer between distributed components, such as edge devices, apps, and backend platforms~\cite{MQTT, 8566599}.
   
\end{itemize}

These protocols complement each other: ISO 15765-4 enables the reliable and real-time retrieval of in-vehicle diagnostics and telemetry via the OBD-II interface; MQTT supports flexible and scalable messaging across heterogeneous devices; the OCPP standardizes station control and monitoring; and ISO 15118 secures direct EV-to-charger interactions. This combination supports seamless connectivity from in-vehicle data acquisition to backend optimization and grid service orchestration.

Furthermore, the private 5G network enhances privacy and operational integrity through network slicing and local spectrum control, isolating critical mobility and charging data from other traffic and external interference. This setup supports near real-time grid services and dynamic tariff response while providing a solid foundation for future city-scale deployments.

\subsection{Design rationale:}
The architecture prioritizes scalability, flexibility, and high responsiveness demanded by urban bidirectional EV charging. Leveraging dedicated 5G connectivity ensures ultra-low latency and reliable communications across all system layers. Its modular design integrates heterogeneous components—wireless vehicle interfaces, edge relays, and cloud-based optimizers—allowing independent development and seamless scaling. The adoption of open, standardized protocols (ISO 15765-4, ISO 15118-20, OCPP, MQTT) ensures interoperability and secure data exchange. User-aware control algorithms balance individual preferences with grid constraints, enabling adaptive, efficient charging. Together, these elements form a robust, future-proof foundation for sustainable urban mobility solutions.

Table~\ref{tab:key-characteristics} summarizes the foundational features of this communication-first architecture, which underpins the system-level interactions detailed next.

\begin{table*}[!t]
\centering
\caption{\mbox{Feature Breakdown of the Communication-First Architecture}}
\label{tab:key-characteristics}
\renewcommand{\arraystretch}{1.2}
\begin{tabular}{|L{5.1cm}|L{12.1cm}|}
\hline
\textbf{Feature} & \textbf{Description} \\
\hline
Pre-connection data exchange & Enables early access to vehicle state (e.g., state-of-charge) and driver preferences prior to plug-in, supporting proactive scheduling and system forecasting. \\
\hline
Modular system design & Flexible integration of heterogeneous components such as OBD-II interfaces, in-car processing unit, tablets, and 5G connectivity, facilitating easy upgrades and vendor independence. \\
\hline
Real-time, bidirectional communication & Continuous exchange of data among user, vehicle, bidirectional charging station, and grid control center, enabling adaptive and coordinated charging management. \\
\hline
Interoperability through open protocols & Utilizes standards like OCPP and wireless interfaces (e.g., Wi-Fi, 5G) to ensure broad compatibility across diverse EV models, charging infrastructure, and network operators. \\
\hline
Scalability for urban deployment & Designed for replication in different urban contexts by leveraging existing wireless and 5G infrastructure, supporting city-specific regulations and mobility ecosystems. \\
\hline
User-aware control logic & Charging strategies dynamically adapt to user inputs (such as desired departure time and energy needs) alongside grid conditions and mobility trends, improving user engagement and system efficiency. \\
\hline
\end{tabular}
\end{table*}

\section{Application Scenarios and Deployment Plan}
\subsection{Real-world context: Dresden Ostra District}
The planned deployment takes place in the Dresden Ostra Sport Park, a sub-site of the Dresden Ostra District, a dynamic urban environment serving as a pilot site within the EU-funded \textit{Mobilities for EU} project. This district plays a key role in Dresden’s Sustainable Urban Mobility Plan~\cite{SUMP2025plus} and is currently being equipped with state-of-the-art bidirectional electric vehicle charging infrastructure. The initiative aligns with both local and European Union objectives for climate neutrality and the development of Positive Energy Districts (PEDs), demonstrating how advanced smart charging and Vehicle-to-Grid (V2G) technologies can transition from laboratory experiments and simulation to practical, scalable urban operations~\cite{SmartCityDresden2025}.

\subsection{Deployment Plan}
The deployment follows a structured, phased approach to ensure comprehensive testing, operational safety, and meaningful stakeholder engagement. In the initial stage, laboratory integration and software-in-the-loop testing are conducted for the wireless OBD-II communication interfaces, in-car processing unit, and the user-facing mobile application within a simulated bidirectional charging environment.

Building on these lab tests, the emulation phase is carried out at the TU Dresden campus parking facilities, where real vehicles equipped with the developed platform interact with the DymoBat optimization system~\cite{dymobat2025}, while simulated user inputs validate logic responsiveness. 

The final phase involves full-scale deployment in the Ostra District. In this phase, bidirectional chargers, interconnected via the city’s private 5G network, are tested under real operating conditions using EVs from the Ostra Sport Park fleet. These EVs are regularly used by facility staff for business-related mobility, providing a realistic context for evaluating system performance in mixed-use, day-to-day urban operations.

\subsection{Key Use Cases}
The architecture supports several representative operational scenarios:

\textit{Pre-connection awareness:} An electric vehicle arriving at a parking location transmits its current SoC and the driver's estimated departure time via the wireless OBD-II interface and mobile app to the central optimizer over a 5G link. This early data enables the control system to forecast available energy resources and schedule charging or discharging operations before the vehicle is physically connected to a charger.

\textit{Personalized charging optimization:} Drivers provide minimal but relevant input via the app, namely, their expected departure time and anticipated driving distance (e.g., short or long trip). These lightweight preferences enable the system to suggest tailored charging strategies that improve both user confidence and energy efficiency.

\textit{Fleet and multi-EV coordination:} The system can scale to environments with multi-vehicle, such as fleet depots. It enables coordinated charging, dynamic load balancing, and prioritization of energy dispatch based on user needs and grid conditions.

\subsection{Anticipated Benefits}
The proposed system offers tangible benefits across multiple stakeholder levels. From a grid management perspective, early access to driver intent and real-time SoC data enables better load forecasting, peak shaving, and operational flexibility, particularly in support of demand response schemes.

From a user perspective, the system demonstrates the feasibility of capturing driver preferences and vehicle telemetry early via wireless communication. Although the current app serves primarily as a minimal proof-of-concept, the communication architecture provides a foundation for future user-centric smart charging services aimed at enhancing transparency and engagement.

Finally, stakeholders such as utilities, municipalities, and city planners benefit from the availability of granular, real-world telemetry collected during field operations. This detailed data supports regulatory compliance, the development of innovative business models, and the planning of resilient, integrated urban energy systems aligned with long-term sustainability goals.

\subsection{Timeline and Milestones}

The pilot rollout is structured in three main stages:
\begin{itemize}
    \item \textbf{Q3 2025:} Completion of laboratory and emulation validation, including early-stage integration of communication modules, app functionality trials, and testing of optimization logic.
    
    \item \textbf{Q4 2025:} Installation of bidirectional charging stations at Ostra Sport Park and initiation of the field pilot using EVs from the facility's operational fleet, driven routinely by staff in their daily duties.
    
    \item \textbf{Late 2025–2026:} Ongoing real-world operation, including data collection, stakeholder engagement, algorithm refinement, and evaluation of technical performance and sustainability impact in alignment with city and EU-level targets.
\end{itemize}

This use-case-driven, phased deployment is designed not only to validate the technical feasibility and scalability of the developed architecture, but also to demonstrate its potential as a foundational element of integrated urban energy systems, supporting the broader goals of the \textit{Mobilities for EU} project.

\section{Discussion}
\textbf{Strengths/advantages:}
This work advances state-of-the-art EV charging systems by moving beyond traditional plug-in and static user settings. Our solution is proactive, enabling early, pre-connection data acquisition and user preference integration, which contrasts with plug-dependent strategies that react only after vehicle connection. The architecture is platform-independent, supporting interoperability across EV models and networks through standard protocols like OCPP. It is also designed to be scalable for wider urban deployment and is user-centric, integrating driver mobility intentions to optimize charging dynamically and transparently.

\textbf{Challenges/risks:} 
Despite these advantages, several challenges remain. Cybersecurity and data privacy risks are inherent when collecting and communicating sensitive user and vehicle telemetry; robust encryption and secure device authentication are essential~\cite{mathew2022data}. Interoperability issues can arise from heterogeneity among EVs, charging station hardware, and network infrastructures, requiring conformance to evolving standards and vendor collaboration~\cite{gupta2025standards}. Furthermore, user adoption may be impacted by trust, usability, and perceived benefits, necessitating clear communication and engagement strategies~\cite{arpaci2025drivers}. On the regulatory front, gaps in legislation and standards around V2G operations within the Dresden and broader EU context may slow implementation and require active policy dialogue~\cite{wang2024b_usecase}.

\textbf{Lessons for wider adoption:}
The communication-first and modular architecture demonstrated here offers a flexible foundation adaptable to diverse urban environments and energy services. Cities can leverage existing wireless and 5G networks to enable user-responsive EV charging ecosystems tailored to their specific mobility patterns and grid needs. Key enablers for adoption include establishing interoperable frameworks that connect stakeholders across transportation and energy sectors and fostering early collaboration to ensure alignment of technical, regulatory, and social objectives. This user-inclusive approach facilitates meaningful engagement with drivers, utilities, and planners, accelerating the integration of scalable and sustainable EV charging solutions.

\section{Conclusion and Future Work}
This paper has introduced a modular, communication-first architecture for advanced bidirectional EV charging, demonstrated through a real-world urban deployment. By enabling early access to vehicle data and driver preferences via wireless interfaces and integrating real-time, bidirectional communication over 5G networks, the proposed system addresses core limitations of conventional, plug-in-dependent charging approaches. The architecture supports scalable urban deployment, user-aware control strategies, and interoperability through open standards, offering a practical foundation for responsive, future-oriented energy-mobility systems.

Key contributions of this work include the deployment and validation of a system in which proactive data exchange enables coordinated charging across vehicles, users, infrastructure, and energy systems. The modular and open design facilitates scenario-based optimization while creating pathways for user participation and integration with urban planning and grid operations.

Future work will focus on expanding pilot deployments to test the architecture under varying regulatory, infrastructural, and social conditions. Additional efforts will target the refinement of dynamic optimization logic based on real-time inputs, along with continued evaluation of system performance, user experience, and operational sustainability. Active collaboration with city planners, utilities, and industry stakeholders will remain central to aligning technical evolution with practical implementation needs.

\section{Acknowledgement}
This work is funded by the European Union under Grant Agreement No 101139666, MOBILITIES FOR EU. Views and opinions expressed are those of the authors only and do not necessarily reflect those of the European Union or the European Climate, Infrastructure and Environment Executive Agency (CINEA). Neither the European Union nor the granting authority can be held responsible for them. It is supported by  the Economic Affairs and Climate Action (BMWK) under project ID 03EI6082A, DymoBat, the German Research Foundation (DFG) as part of Germany's Excellence Strategy—EXC 2050/1—Cluster of Excellence “Centre for Tactile Internet with Human-in-the-Loop” (CeTI) of Technische Universität Dresden under project ID 390696704 and the Federal Ministry of Education and Research (BMBF) in the program of “Souverän. Digital. Vernetzt.” Joint project 6G-life, grant number 16KISK001K.

\section{Reference}
\bibliographystyle{IEEEtran}
\bibliography{references}

@article{prata2013moving,
  title={Moving towards the sustainable city: The role of electric vehicles, renewable energy and energy efficiency},
  author={Prata, J and Arsenio, E and Pontes, JP},
  journal={WIT Transactions on Ecology and the Environment},
  volume={179},
  pages={871--883},
  year={2013},
  publisher={WIT Press}
}

@INPROCEEDINGS{wang2024b_usecase,
  AUTHOR = "S. Wang and J. A. Cabrera and F. H. P. Fitzek",
  TITLE = "Bidirectional Charging Use Cases: Innovations in E-Mobility and Power-Grid Flexibility",
  BOOKTITLE = "IEEE International Smart Cities Conference (ISC2)",
  YEAR = "2024",
  pages = "1--6",
  address = "Pattaya, Thailand"
}

@misc{MobilitiesForEU,
  author = {MobilitiesforEU},
  title = {Mobilities for EU},
  year = {2024},
  howpublished = {\url{https://mobilities-for.eu/}},
  note = {EU Horizon Project},
}

@INPROCEEDINGS{wang2024a5G,
  AUTHOR = "S. Wang and C. Lehmann and R. Radeke and F. H. P. Fitzek",
  TITLE = "Enabling Sustainable Urban Mobility: The Role of 5G Communication in the Mobilities for EU Project",
  BOOKTITLE = "IEEE International Smart Cities Conference (ISC2)",
  YEAR = "2024",
  pages = "1--6",
  address = "Pattaya, Thailand"
}

@article{delRioCarazo2022,
  author    = {Laura del Río Carazo and Emiliano Acquila-Natale and Santiago Iglesias-Pradas and Ángel Hernández-García},
  title     = {Sustainable Rural Electrification Project Management: An Analysis of Three Case Studies},
  journal   = {Energies},
  year      = {2022},
  volume    = {15},
  number    = {3},
  pages     = {1203},
  doi       = {10.3390/en15031203}
}

@inproceedings{Wang2025:Simulationb,
title = {Simulation Architecture for Electric Vehicle Charging Optimization in Dresden\'s Ostra District},
author = {Shangqing {Wang} and Syed Irtaza {Haider} and Shiwei {Shen} and Faezeh {Motazedian} and Rico {Radeke} and Frank H. P. {Fitzek}},
doi = {10.5220/0013445000003953},
issn = {2184-4968},
year = {2025},
date = {2025-01-01},
urldate = {2025-01-01},
booktitle = {International Conference on Smart Cities and Green ICT Systems (SMARTGREENS)},
pages = {56\textendash65},
pubstate = {published},
tppubtype = {inproceedings}
}

@techreport{CIVITAS2017,
  author       = {Dirk Engels and Gitte Van Den Bergh and Tim Breemersch},
  title        = {Refined CIVITAS Process and Impact Evaluation Framework},
  institution  = {CIVITAS SATELLITE Project, European Commission},
  year         = {2017},
  number       = {D2.3},
  note         = {Deliverable No.: D2.3, Grant Agreement No.: 713813, Work package No.: WP2},
  howpublished = {Public deliverable},
  address      = {Brussels, Belgium},
  month        = {August},
}

@misc{SUMP2025plus,
  author       = {{City of Dresden, Urban Development Division}},
  title        = {Sustainable Urban Mobility Plan 2025plus: An Overview},
  year         = {2016},
  month        = {June},
  address      = {Dresden, Germany},
}

@article{micari2024electric,
  title={Electric Vehicles for a flexible energy system: Challenges and opportunities},
  author={Micari, Salvatore and Napoli, Giuseppe},
  journal={Energies},
  volume={17},
  number={22},
  pages={5614},
  year={2024},
  publisher={MDPI}
}

@inproceedings{rehman2019cloud,
  title={A cloud-based development environment using HLA and kubernetes for the co-simulation of a corporate electric vehicle fleet},
  author={Rehman, Kasim and Kipouridis, Orthodoxos and Karnouskos, Stamatis and Frendo, Oliver and Dickel, Helge and Lipps, Jonas and Verzano, Nemrude},
  booktitle={2019 IEEE/SICE International Symposium on System Integration (SII)},
  pages={47--54},
  year={2019},
  organization={IEEE}
}

@article{IEEE_HLA_2010,
  author    = {},
  title     = {{IEEE} {S}tandard for {M}odeling and {S}imulation ({M\&S}) High Level Architecture ({HLA})-- Object Model Template (OMT) Specification}, 
  journal   = {IEEE Std 1516.2-2010 (Revision of IEEE Std 1516.2-2000)}, 
  year      = {2010},
  doi       = {10.1109/IEEESTD.2010.5557731}}

@inproceedings{haumer1999bridging,
  title={Bridging the gap between past and future in RE: a scenario-based approach},
  author={Haumer, Peter and Heymans, Patrick and Jarke, Matthias and Pohl, Klaus},
  booktitle={Proceedings IEEE International Symposium on Requirements Engineering (Cat. No. PR00188)},
  pages={66--73},
  year={1999},
  organization={IEEE}
}

@unpublished{Wang2025:PDTT,
  author    = {Shangqing Wang and Syed Irtaza Haider and Shiwei Shen and Faezeh Motazedian and Rico Radeke and Frank H. P. Fitzek},
  title     = {From Simulation to Sustainable Urban Mobility: A Purpose-Driven Framework for Advanced EV Charging Technology Transfer},
  note      = {Invited paper, submitted to Proceedings of SMARTGREENS 2025 (Porto, Portugal: Springer LNCS/LNAI, 2025)},
  year      = {2025}
}

@article{wu2018transactive,
  title={Transactive real-time electric vehicle charging management for commercial buildings with PV on-site generation},
  author={Wu, Qiuwei and Shahidehpour, Mohammad and Li, Canbing and Huang, Shaojun and Wei, Wei and others},
  journal={IEEE Transactions on Smart Grid},
  volume={10},
  number={5},
  pages={4939--4950},
  year={2018},
  publisher={IEEE}
}

@article{Schellenberg2021telematics,
 author = {Schellenberg, C. and Filler, T. and et al.},
 title = {Remote electric vehicle battery monitoring and state of health estimation using telematics data},
 journal = {IEEE Transactions on Vehicular Technology},
 year = {2021}
}

@article{Nordholm2023EVreview,
 author = {Nordholm, J. and Westergren, K. and Johansson, R.},
 title = {Integration of Electric Vehicles into Power Grids—A Review of Commercial and Research Solutions for EV Fleet Management},
 journal = {Energies},
 year = {2023}
}

@misc{SciurusFinalReport,
author = {Catapult, Energy Systems},
title = {Project Sciurus Final Report},
year = {2021},
note = {\url{https://es.catapult.org.uk/news/project-sciurus-final-report/}}
}

@misc{ParkerProject,
author = {Parker Project Consortium},
title = {Parker Project: Final Results and Recommendations},
year = {2020},
note = {\url{https://parker-project.com/}}
}

@misc{dymobat2025,
    title = {DymoBat Research Project},
    note = {Available at \url{https://dymobat.de/}, Accessed: 22 July 2025},
    year = {2025}
}

@misc{SmartCityDresden2025,
  author = {{Smart City Dresden}},
  title = {Mobilities for EU - Projektübersicht Dresden},
  year = {2025},
  howpublished = {\url{https://smartcity.dresden.de/projekte/mobilities-for-eu}},
  note = {Accessed: 22 July 2025}
}

@article{mathew2022data,
  title={Data privacy and security concerns in IoT-based traffic surveillance},
  author={Mathew, Bamidele},
  year={2022}
}

@incollection{gupta2025standards,
  title={Standards and Protocols for Interoperability in Smart EV Systems},
  author={Gupta, Sandeep and Varshney, Tarun and Saini, Parvesh and Hussain, Muhammad Majid},
  booktitle={Modern Computing Technologies for EV Efficiency and Sustainable Energy Integration},
  pages={115--150},
  year={2025},
  publisher={IGI Global Scientific Publishing}
}

@article{arpaci2025drivers,
  title={Drivers of vehicle-to-everything (V2X) adoption: A behavioral reasoning theory perspective},
  author={Arpaci, Ibrahim and Al-Sharafi, Mohammed A and Mahmoud, Moamin A},
  journal={PloS one},
  volume={20},
  number={7},
  pages={e0327084},
  year={2025},
  publisher={Public Library of Science San Francisco, CA USA}
}

@standard{ISO15118-20,
  title        = {ISO 15118-20: Road vehicles — Vehicle to grid communication interface — Part 20: 2nd generation network layer and application layer requirements},
  organization = {International Organization for Standardization},
  year         = {2022},
  url          = {https://www.iso.org/standard/77845.html}
}

@misc{OCPP,
  title        = {Open Charge Point Protocol (OCPP)},
  organization = {Open Charge Alliance},
  year         = {2024},
  url          = {https://openchargealliance.org/protocols/open-charge-point-protocol/}
}

@misc{MQTT,
  title        = {Message Queuing Telemetry Transport (MQTT)},
  organization = {OASIS},
  year         = {2019},
  url          = {https://mqtt.org/}
}

@standard{ISO15765-4,
  title        = {ISO 15765-4: Road vehicles — Diagnostic communication over Controller Area Network (DoCAN) — Part 4: Requirements for emissions-related systems},
  organization = {International Organization for Standardization},
  year         = {2021},
  url          = {https://www.iso.org/standard/78384.html}
}

@INPROCEEDINGS{8566599,
  author={Yerlikaya, {\"O}zlem and Dalk{\i}l{\i}c, G{\"o}khan},
  booktitle={2018 3rd International Conference on Computer Science and Engineering (UBMK)}, 
  title={Authentication and Authorization Mechanism on Message Queue Telemetry Transport Protocol}, 
  year={2018},
  volume={},
  number={},
  pages={145-150},
  doi={10.1109/UBMK.2018.8566599}}
\end{document}